\pdfoutput=1

\newif\iffull
\fullfalse

\documentclass[letterpaper,twocolumn,10pt]{article}
\usepackage{usenix,epsfig,endnotes}

\usepackage{amssymb}
\usepackage[all]{xy}
\usepackage{url}
\usepackage{tikz}
\usetikzlibrary{automata,arrows,positioning,arrows.meta,calc}
\usepackage[inline]{enumitem}
\usepackage{listings}
\usepackage{multirow}
\usepackage{xifthen}
\usepackage{amsmath}
\usepackage{xspace}
\usepackage{booktabs}
\usepackage{balance}
\usepackage{pifont}
\usepackage{color}
\usepackage{todonotes}

\usepackage{infer}
\usepackage{mathtools}
\usepackage{xargs}
\usepackage{xfrac}
\usepackage{amsthm}
\usepackage{ stmaryrd }

\newcounter{ncomm}

\let\state\relax 
\newcommand{\cf}{\emph{cf.}}
\newcommand{\eg}{\emph{e.g.}}
\newcommand{\ie}{\emph{i.e.}}
\newcommand{\etal}{\emph{et al.}}

\newcommand{\WPE}{\ensuremath{\text{WPSE}}\xspace}

\newcommand{\binder}[2][]{\textbf{#2}\ifthenelse{\isempty{#1}}{}{\textbf{\{}#1\textbf{\}}}\textbf{:=}}
\newcommand{\binderx}[3][]{\textbf{#2}\ifthenelse{\isempty{#1}}{}{\textbf{\{}#1\textbf{\}}}\textbf{:=}\big[{#3}\big]}
\newcommand{\usage}[1]{\texttt{\textbf{#1}}}

\newcommand{\user}{U}
\newcommand{\rp}{\mathit{RP}}
\newcommand{\idp}{\mathit{IdP}}
\newcommand{\sep}{\mathit{SP}}
\newcommand{\webbrowser}{\mathit{C}}
\newcommand{\params}[1]{{\small \textsf{#1}}}

\newcommand{\aidp}{\mathit{AIdP}}
\newcommand{\hidp}{\mathit{HIdP}}
\newcommand{\circled}[1]{\raisebox{.5pt}{\textcircled{\raisebox{-.9pt} {#1}}}}
\newcommand{\ident}[1]{\texttt{\textbf{#1}}}

\newcommand{\reqshape}[2]{{#1}\langle {#2} \rangle}
\newcommand{\respshape}[2]{{#1}({#2})}
\newcommand{\state}[1]{\textit{#1}}

\newtheorem{theorem}{Theorem}
\newcommand{\network}{N}
\newcommand{\browser}{B}
\newcommand{\monitor}{M}
\newcommand{\orig}{S_{\textit{orig}}}
\newcommand{\monitored}{S_\textit{new}}
\newcommand{\app}{\textit{App}}
\newcommand{\projected}[2]{{#1} \downarrow {#2}}
\newcommand{\subsumes}{\preccurlyeq}
\newcommand*\circledbig[1]{\tikz[baseline=(char.base)]{
            \node[shape=circle,draw,inner sep=1pt] (char) {#1};}}

\newcommand{\conc}{\circledbig{\textit{C}}}
\newcommand{\pipar}[2]{#1| #2}
\newcommand{\piin}[3]{{#1}(#2). {#3}}
\newcommand{\piout}[3]{\overline{#1}\langle {#2} \rangle. {#3}}
\newcommand{\pirep}[1]{!{#1}}
\newcommand{\pistop}{0}
\newcommand{\pires}[2]{v{#1}.{#2}}
\newcommand{\piif}[3]{\textit{if}~{#1}~\textit{then}~{#2}~\textit{else}~{#3}} % maybe we don't need this 
\newcommand{\expar}[2]{#1 | #2}
\newcommand{\exnres}[2]{v{#1}.{#2}}
\newcommand{\exvres}[2]{v{#1}.{#2}}
\newcommand{\exsubs}[2]{\{ \sfrac{#1}{#2} \}}

\newcommand{\Vars}{\mathcal{V}}
\newcommand{\Names}{\mathcal{N}}
\newcommand{\Terms}{\mathcal{T}}
\newcommand{\pframe}[1]{\varphi({#1})}
\newcommand{\domain}[1]{\textit{dom}({#1})}
\newcommand{\inlabel}[2]{{#1}(#2)}
\newcommand{\outlabel}[2]{#1 \langle #2 \rangle}
\newcommand{\reslabel}[2]{v {#2}.\overline{#1} \langle {#2} \rangle}

\newcommand{\transformer}[3]{(#1)_{#2 \leadsto #3}}

\newcommand{\intred}[2]{{#1} \rightarrow {#2}}
\newcommand{\subst}[3]{#1\{\sfrac{#2}{#3} \}}
\newcommand{\eqterms}{=_{E}}
\newcommand{\tracered}[3]{{#1} \xrightarrow{{#3}} {#2}}
\newcommand{\labred}[3]{{#1} \xrightarrow{#3} {#2}}
\newcommand{\traceeq}{\approx_{T}}
\newcommand{\tracesub}{\preccurlyeq_{T}}
\newcommand{\ptrace}[2]{{#1}\,\downharpoonright \, {#2}}
\newcommand{\traceproj}[2]{{#1} \downarrow {#2}}
\newcommand{\concretetraces}[1]{\mathcal{T}_{C}(#1)}
\newcommand{\applysubst}[2]{#1 (#2)}
\newcommand{\ctrace}{\pi_{C}}

\newcommand{\bappin}{ba_{\textit{in}}}
\newcommand{\bappout}{ba_{\textit{out}}}
\newcommand{\bservin}{bs_\textit{in}}
\newcommand{\bservout}{bs_\textit{out}}
\newcommand{\browserp}[4]{B{\scriptstyle(#1, #2, #3, #4)}}
\newcommand{\appp}[2]{\textit{App}{\scriptstyle(#1, #2)}}
\newcommand{\networkp}[3]{N{\scriptstyle(#1, #2, #3)}}
\newcommand{\monitorp}[4]{M{\scriptstyle(#1, #2, #3, #4)}}
\newcommand{\secrets}{\widetilde{s}}

\newcommand{\freenames}[1]{\textit{fn}(#1)}
\newcommand{\freevariables}[1]{\textit{fv}(#1)}

\iffull
\newtheorem{definition}{Definition}
\newtheorem{lemma}{Lemma}

\newcommand{\NN}{\mathbb{N}}
\newcommand{\BB}{\mathbb{B}}
\renewcommand{\pipar}[2]{#1| #2}
\renewcommand{\piin}[3]{{#1}(#2). {#3}}
\renewcommand{\piout}[3]{\overline{#1}\langle {#2} \rangle. {#3}}
\renewcommand{\pirep}[1]{!{#1}}
\renewcommand{\pistop}{0}
\renewcommand{\pires}[2]{v{#1}.{#2}}
\renewcommand{\piif}[3]{\textit{if}~{#1}~\textit{then}~{#2}~\textit{else}~{#3}} % maybe we don't need this 
\renewcommand{\expar}[2]{#1 | #2}
\renewcommand{\exnres}[2]{v{#1}.{#2}}
\renewcommand{\exvres}[2]{v{#1}.{#2}}
\renewcommand{\exsubs}[2]{\{ \sfrac{#1}{#2} \}}

\renewcommand{\Vars}{\mathcal{V}}
\renewcommand{\Names}{\mathcal{N}}
\newcommand{\channelnames}{\mathcal{C}}
\renewcommand{\Terms}{\mathcal{T}}
\renewcommand{\pframe}[1]{\varphi({#1})}
\renewcommand{\domain}[1]{\textit{dom}({#1})}
\renewcommand{\inlabel}[2]{{#1}(#2)}
\renewcommand{\outlabel}[2]{#1 \langle #2 \rangle}
\renewcommand{\reslabel}[2]{v {#2}.\overline{#1} \langle {#2} \rangle}

\renewcommand{\transformer}[3]{(#1)_{#2 \leadsto #3}}
\newcommand{\actions}{\mathcal{A}}
\newcommand{\sequenceof}[1]{\mathcal{S}(#1)}

\renewcommand{\intred}[2]{{#1} \rightarrow {#2}}
\renewcommand{\subst}[3]{#1\{\sfrac{#2}{#3} \}}
\renewcommand{\eqterms}{=_{E}}
\renewcommand{\tracered}[3]{{#1} \xrightarrow{{#3}} {#2}}
\renewcommand{\labred}[3]{{#1} \xrightarrow{#3} {#2}}
\renewcommand{\traceeq}{\approx_{T}}
\renewcommand{\tracesub}{\preccurlyeq_{T}}
\newcommand{\tracesubsume}[1]{\preccurlyeq_{#1}}

\renewcommand{\ptrace}[2]{{#1}~\lightning~ {#2}}
\renewcommand{\traceproj}[2]{{#1} \downarrow_{#2}}
\renewcommand{\concretetraces}[1]{\mathcal{T}_{C}(#1)}
\renewcommand{\applysubst}[2]{#1 (#2)}
\renewcommand{\ctrace}{\pi_{C}}

\newcommand{\tfilter}[2]{{#1} \downarrow_{#2}}

\renewcommand{\bappin}{a_{\textit{i}}}
\renewcommand{\bappout}{a_{\textit{o}}}
\renewcommand{\bservin}{s_\textit{i}}
\renewcommand{\bservout}{s_\textit{o}}
\newcommand{\buserin}{u_i}
\renewcommand{\browserp}[5]{B(#1, #2, #3, #4, #5)}
\renewcommand{\appp}[2]{\textit{App}(#1, #2)}
\renewcommand{\networkp}[3]{N(#1, #2, #3)}
\renewcommand{\monitorp}[5]{M(#1, #2, #3, #4, #5)}
\newcommand{\userp}[2]{U(#1, #2)}
\renewcommand{\secrets}{\widetilde{s}}

\newcommand{\mon}{\monitorp{\bappin}{\bappout}{\bservin}{\bservout}{\buserin}}
\newcommand{\net}{\networkp{\bservin}{\bservout}{\secrets}}
\newcommand{\brow}{\browserp{\bappin}{\bappout}{\bservin}{\bservout}{\buserin}}
\renewcommand{\app}{\appp{\bappin}{\bappout}}

\renewcommand{\user}{\userp{\buserin}{\secrets}}
\newcommand{\full}{\exnres{\bappout, \buserin}{\exvres{\secrets}{\pipar{\transformer{\appp{\bappin}{\bappout}}{\bappout}{e}}{\pipar{\monitorp{\bappin}{\bappout}{\bservin}{\bservout}}{\pipar{\userp{\buserin}{\secrets}}{\networkp{\bservin}{\bservout}{\secrets}}}}}}}

\renewcommand{\freenames}[1]{\textit{fn}(#1)}
\renewcommand{\freevariables}[1]{\textit{fv}(#1)}
\newcommand{\boundnames}[1]{\textit{bn}(#1)}
\newcommand{\boundvariables}[1]{\textit{bv}(#1)}
\newcommand{\freechannels}[1]{\textit{fc}(#1)}

\newcommand{\corres}[4]{\overline{#1}\langle #2 \rangle \leadsto \overline{#3} \langle #4 \rangle}
\fi

\newlist{inlinelist}{enumerate*}{4}
\setlist[inlinelist]{label=\em\roman*)}

\title{\WPE: Fortifying Web Protocols via Browser-Side Security Monitoring}
\author{Stefano Calzavara \\
Universit\`a Ca' Foscari Venezia \\
calzavara@dais.unive.it
\and
Riccardo Focardi \\
Universit\`a Ca' Foscari Venezia \\
focardi@unive.it
\and
Matteo Maffei \\
TU Wien \\
matteo.maffei@tuwien.ac.at
\and
Clara Schneidewind \\
TU Wien \\
clara.schneidewind@tuwien.ac.at
\and
Marco Squarcina \\
Universit\`a Ca' Foscari Venezia \\
squarcina@unive.it \\
\and
Mauro Tempesta \\
Universit\`a Ca' Foscari Venezia \\
tempesta@unive.it}

\date{} % don't want date printed

\begin{document}
\iffull
\onecolumn 
\fi
\maketitle

\subsection*{Abstract}
We present \WPE, a browser-side security monitor for web protocols designed to ensure compliance with the intended protocol flow, as well as confidentiality and integrity properties of messages. We formally prove that \WPE is expressive enough to protect web applications from a wide range of protocol implementation bugs and web attacks. We discuss concrete examples of attacks which can be prevented by \WPE on OAuth 2.0 and SAML 2.0, including a novel attack on the Google implementation of SAML 2.0 which we discovered by formalizing the protocol specification in \WPE. Moreover, we use \WPE to carry out an extensive experimental evaluation of OAuth 2.0 in the wild. Out of 90 tested websites, we identify security flaws in 55 websites (61.1\%), including new critical vulnerabilities introduced by tracking libraries such as Facebook Pixel, all of which fixable by \WPE{}. Finally, we show that \WPE{} works flawlessly on 83 websites (92.2\%), with the 7 compatibility issues being caused by custom implementations deviating from the OAuth 2.0 specification, one of which introducing a critical vulnerability.

\noindent\makebox[\linewidth]{\rule{\columnwidth}{0.4pt}}
This is a preprint of an article accepted for publication in the proceedings of USENIX Security '18.

\section{Introduction}
% !TeX root = main.tex
Web protocols are security protocols deployed on top of HTTP and HTTPS, most notably to implement authentication and authorization at remote servers. Popular examples of web protocols include OAuth 2.0, OpenID Connect, SAML 2.0 and Shibboleth, which are routinely used by millions of users to access security-sensitive functionalities on their personal accounts.
 
Unfortunately,  designing and implementing web protocols is a particular error-prone task even for security experts, as witnessed by the large number of vulnerabilities reported in the literature~\cite{SunB12,BansalBDM14,BansalBDM13,ZhouE14,LiM16,LiM14,YangLLZH16,WangCW12}. The main reason for this is that web protocols involve communication with a web browser, which does not strictly follow the protocol specification, but reacts asynchronously to any input it receives, producing messages which may have an import on protocol security. Reactiveness is dangerous because the browser is agnostic to the web protocol semantics: it does not know when the protocol starts, nor when it ends, and is unaware of the order in which messages should be processed, as well as of the confidentiality and integrity guarantees desired for a protocol run.  For example, in the context of OAuth 2.0, Bansal \etal~\cite{BansalBDM14} discussed \emph{token redirection attacks} enabled by the presence of open redirectors, while Fett \etal~\cite{FettKS16} presented \emph{state leak attacks} enabled by the communication of the Referer header; these attacks are not apparent from the protocol specification alone, but come from the subtleties of the browser behaviour. 

Major service providers try to aid software developers to correctly integrate web protocols in their websites by means of JavaScript APIs; however, web developers are not forced to use them, can still use them incorrectly~\cite{WZCQEG13}, and the APIs themselves do not necessarily implement the best security practices~\cite{SunB12}. This unfortunate situation led to the proliferation of attacks against web protocols even at popular services.

In this paper, we propose a fundamental paradigm shift to strengthen the security guarantees of web protocols. The key idea we put forward is to extend browsers with a security monitor which is able to enforce the compliance of browser behaviours with respect to the web protocol specification. This approach brings two main benefits:
\begin{enumerate}
\item web applications are automatically protected against a large class of  bugs and vulnerabilities on the browser-side,  since the browser is aware of the intended protocol flow and any deviation from it is detected at runtime;
\item protocol specifications can be written and verified once, possibly as a community effort, and then uniformly enforced at a number of different websites by the browser.
\end{enumerate}

Remarkably, though changing the behaviour of web browsers is always delicate for backward compatibility, the security monitor we propose is carefully designed to interact gracefully with existing websites, so that the website functionality is preserved unless it critically deviates from the intended protocol specification. Moreover, a large set of the monitor functionalities can be implemented as a browser extension, thereby offering immediate protection to Internet users and promising a significant practical impact.

\subsection{Contributions}
In this paper, we make the following contributions:
\begin{enumerate}
	\item we identify three fundamental browser-side security properties  for  web protocols, that is, the  \emph{confidentiality} and \emph{integrity} of message components, as well as the compliance with the intended \emph{protocol flow}. We discuss concrete examples of their import for the popular authorization protocol OAuth 2.0;
	\item we semantically characterize these properties and formally prove that their enforcement  suffices  to protect the web application from a wide range of protocol implementation bugs and attacks on the application code running in the browser; 
	\item we propose the Web Protocol Security Enforcer, or \WPE{} for short, a browser-side security monitor designed to enforce the aforementioned security properties,  which we implement as a publicly available Google Chrome extension; 
	\item we experimentally assess the effectiveness of \WPE{} by testing it against 90 popular websites making use of OAuth 2.0 to implement single sign-on at major identity providers. In our analysis, we identified security flaws in 55 websites (61.1\%), including new critical vulnerabilities caused by tracking libraries such as Facebook Pixel, all of which fixable by \WPE{}. We show that WPSE works flawlessly on 83 websites (92.2\%), with the 7 compatibility issues being caused by custom implementations deviating from the OAuth 2.0 specification, one of which introducing a critical vulnerability;
	\item to show the generality of our approach, we also considered SAML 2.0, a popular web authorization protocol: while formalizing its specification, we found a new attack on the Google implementation of SAML 2.0 that has been awarded a bug bounty according to the Google Vulnerability Reward Program.\footnote{\hspace{4pt}\url{https://www.google.com/about/appsecurity/reward-program/}}
\end{enumerate}

\section{Security Challenges in Web Protocols}
\label{sec:challenges}
% !TeX root = main.tex

The design of web protocols comes with various security challenges which can often be attributed to the presence of the web browser that acts as a non-standard protocol participant. In the following, we discuss three crucial challenges, using the OAuth 2.0 authorization protocol as illustrative example. 

\subsection{Background on OAuth 2.0}
OAuth 2.0~\cite{RFC6749} is a web protocol that enables resource owners to grant controlled access to resources hosted at remote servers. Typically, OAuth 2.0 is also used for authenticating the resource owner to third parties by giving them access to the resource owner's identity stored at an identity provider. This functionality is known as Single Sign-On (SSO). Using standard terminology, we refer to the third-party application as \emph{relying party} ($\rp$) and to the website storing the resources, including the identity, as \emph{identity provider} ($\idp$).\footnote{~The OAuth 2.0 specification distinguishes between \emph{resource servers} and \emph{authorization servers} instead of considering one identity provider that stores the user's identity as well as its resources~\cite{RFC6749}, but it is common practice to unify resource and authorization servers as one party~\cite{FettKS16,SunB12,LiM14}.}

The OAuth 2.0 specification defines four different protocol flows, also known as \textit{grant types} or \emph{modes}. We focus on the \emph{authorization code} mode and the \emph{implicit} mode since they are the most commonly used by websites.

\tikzset{>={To[length=2.5mm,width=1.5mm]}}
\begin{figure*}[htb]
	\centering
	\def\hd{0.6cm}
	\begin{tikzpicture}[node distance=7cm and 1cm,  font=\ttfamily\footnotesize]
		\node(user){$\user$};
		\node[right of = user](rp){$\rp$}; 
		\node[right of = rp](idp){$\idp$}; 
		\node(rpredirect) at ($(rp)-(0, 7*\hd)$) {$\rp(\textit{redirect URI})$};
		
		\path[draw, ->] ($(user)-(0, \hd)$) to node[above]{\circled{1} $\idp$} ($(rp)-(0, \hd)$); 
		\path[draw, -] ($(user)-(0, 2*\hd)$) to node[above]{} ($(rp)-(0, 2*\hd)$); 
		\path[draw, -] ($(user)-(0, 2*\hd)$) to ($(user)-(0, 3*\hd)$); 
		\path[draw, ->] ($(user)-(0, 3*\hd)$) to node[above]{\circled{2} $\rp$ ID, redirect URI, state} ($(idp)-(0, 3*\hd)$); 
		\path[draw, <-] ($(user)-(0, 4*\hd)$) to node[above]{\circled{3} Login form} ($(idp)-(0, 4*\hd)$); 
		\path[draw, ->] ($(user)-(0, 5*\hd)$) to node[above]{User credentials} ($(idp)-(0, 5*\hd)$); 
		\path[draw, -] ($(user)-(0, 6*\hd)$) to node[above]{} ($(idp)-(0, 6*\hd)$); 
		\path[draw, -] ($(user)-(0, 6*\hd)$) to ($(user)-(0, 7*\hd)$); 
		\path[draw, ->] ($(user)-(0, 7*\hd)$) to node[above]{\circled{4} authorization code, state} (rpredirect); 
		\path[draw, ->] ($(rp)-(0, 8*\hd)$) to node[above]{\circled{5} authorization code, $\rp$ ID, redirect URI} ($(idp)-(0, 8*\hd)$); 
		\path[draw, <-] ($(rp)-(0, 9*\hd)$) to node[above]{\circled{6} access token} ($(idp)-(0, 9*\hd)$); 
		\path[draw, ->] ($(rp)-(0, 10*\hd)$) to node[above]{\circled{7} access token} ($(idp)-(0, 10*\hd)$); 
		\path[draw, <-] ($(rp)-(0, 11*\hd)$) to node[above]{\circled{8} resource} ($(idp)-(0, 11*\hd)$); 
	\end{tikzpicture}
	\caption{OAuth 2.0 (authorization code mode).}
	\label{fig:oauth-explicit}
\end{figure*}
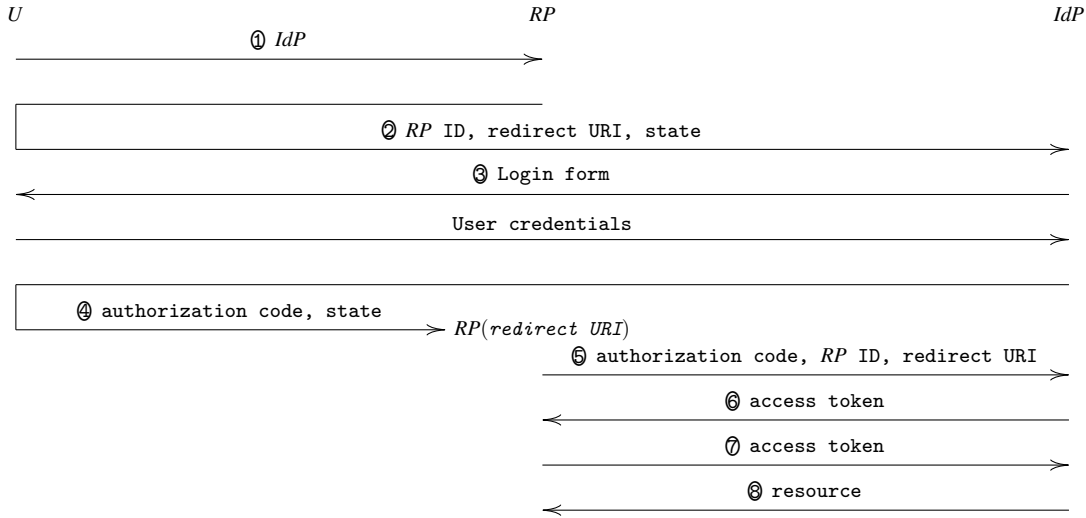

The authorization code mode is intended for a $\rp$ whose main functionality is carried out at the server side. The high-level protocol flow is depicted in Figure~\ref{fig:oauth-explicit}. For the sake of readability, we introduce a simplified version of the protocol abstracting from some implementation details that are presented in Section~\ref{subsec:oauth}. The protocol works as follows:
\begin{enumerate}[label=\protect\circled{\arabic*}]
\item the user $\user$ sends a request to $\rp$ for accessing a remote resource. The request specifies the $\idp$ that holds the resource. In the case of SSO, this step determines which $\idp$ should be used;
\item $\rp$ redirects $\user$ to the login endpoint of $\idp$. This request contains the $\rp$'s identity at $\idp$, the URI that $\idp$ should redirect to after successful login and an optional state parameter for CSRF protection that should be bound to $\user$'s state;
\item $\idp$ answers to the authorization request with a login form and the user provides her credentials;
\item $\idp$ redirects $\user$ to the URI of $\rp$ specified at step \circled{2}, including the previously received state parameter and an authorization code;
\item $\rp$ makes a request to $\idp$ with the authorization code, including its identity, the redirect URI and optionally a shared secret with the $\idp$;
\item $\idp$ answers with an access token to $\rp$;
\item $\rp$ makes a request for the user's resource to $\idp$, including the access token;
\item $\idp$ answers $\rp$ with the user's resource at $\idp$.
\end{enumerate}
The implicit mode differs from the authorization code mode in steps \circled{4}-\circled{6}. Instead of granting an authorization code to $\rp$, the $\idp$ provides an access token in the fragment identifier of the redirect URI. A piece of JavaScript code embedded in the page located at the redirect URI extracts the access token and communicates it to the $\rp$.

\subsection{Challenge \#1: Protocol Flow}
\label{subsec:sessionswapping}
Protocols are specified in terms of a number of sequential message exchanges which honest participants are expected to follow, but the browser is not forced to comply with the intended protocol flow.

\textbf{Example in OAuth 2.0.}
The use of the state parameter is recommended to prevent attacks leveraging this idiosyncrasy. When OAuth is used to implement SSO and $\rp$ does not provide the state parameter in its authorization request to $\idp$ at step \circled{2}, it is possible to force the honest user's browser to authenticate as the attacker. This attack is known as \emph{session swapping}~\cite{SunB12}.

We give a short overview on this attack against the authorization code mode. A web attacker $A$ initiates SSO at $\rp$ with an identity provider $\idp$, performs steps \circled{1}-\circled{3} of the protocol and learns a valid authorization code for her session. Next, $A$ creates a page on her website that, when visited, automatically triggers a request to the redirect URI of $\rp$ and includes the authorization code.
When a honest user visits this page, the login procedure is completed at $\rp$ and an attacker session is established in the user's browser.
 
\subsection{Challenge \#2: Secrecy of Messages}
\label{subsec:stateleak}
The security of protocols typically relies on the confidentiality of cryptographic keys and credentials, but the browser is not aware of which data must be kept secret for protocol security.

\textbf{Example in OAuth 2.0.}
The secrecy of the authorization credentials (namely authorization codes and access tokens) is crucial for meeting the protocol security requirements, since their knowledge allows an attacker to access the user's resources. The secrecy of the state parameter is also important to ensure session integrity. 

An example of an unintended secrets leakage is the \emph{state leak} attack described in~\cite{FettKS16}. If the page loaded at the redirect URI in step \circled{4} loads a resource from a malicious server, the state parameter and the authorization code (that are part of the URL) are leaked in the Referer header of the outgoing request. The learned authorization code can potentially be used to obtain a valid access token for $\user$ at $\idp$, while the leaked state parameter enables the session swapping attack discussed previously.

\subsection{Challenge \#3: Integrity of Messages}
\label{subsec:naivesessionintegrity}
Protocol participants are typically expected to perform a number of runtime checks to prove the integrity of the messages they receive and ensure the integrity of the messages they send, but the browser cannot perform these checks unless they are explicitly carried out in a JavaScript implementation of the web protocol.

\textbf{Example in OAuth 2.0.}
An attack that exploits this weakness is the \emph{na\"ive $\rp$ session integrity} attack presented in~\cite{FettKS16}. Suppose that $\rp$ supports SSO with various identity providers and uses different redirect URIs to distinguish between them. In this case, an attacker controlling a malicious identity provider $\aidp$ can confuse the $\rp$ about which provider is being used and force the user's browser to login as the attacker. 

To this end, the attacker starts a SSO login at $\rp$ with an honest identity provider $\hidp$ to obtain a valid authorization code for her account. If a honest user starts a login procedure at $\rp$ with $\aidp$, in step \circled{4} $\aidp$ is expected to redirect the user to $\aidp$'s redirect URI at $\rp$. If $\aidp$ redirects to the redirect URI of $\hidp$ with the authorization code from the attacker session, then $\rp$ mistakenly assumes that the user intended to login with $\hidp$. Therefore, $\rp$ completes the login with $\hidp$ using the attacker's account. 

\section{\WPE: Design and Implementation}
% !TeX root = main.tex

The Web Protocol Security Enforcer (\WPE) is the first browser-side security monitor addressing the peculiar challenges of web protocols. The current prototype is implemented as an extension for Google Chrome, which we make available online.\footnote{~\url{https://sites.google.com/site/wpseproject/}}

\subsection{Key Ideas of \WPE}
We illustrate \WPE{} on the authorization code mode of OAuth 2.0, where Google is used as identity provider and the state parameter is not used (since it is not mandatory at Google). For simplicity, here we show only the most common scenario where the user has an ongoing session with the identity provider and the authorization to access the user's resources on the provider has been previously granted to the relying party.

\vspace{-5pt}
\subsubsection{Protocol Flow}

\begin{figure*}[htb]
	\centering
	\begin{tikzpicture}[->,auto, node distance=3.2cm,semithick]
	  \tikzstyle{every state}=[fill=white,draw=black,text=black]
	
	  \node[initial,state,minimum size=1.2cm]     (A)              {$\mathit{init}$};
	  \node[state,minimum size=1.2cm]             (B) [right of=A] {$\mathit{auth}$};
	  \node[state,minimum size=1.2cm]             (C) [right of=B] {$\mathit{access}$};
	  \node[state, accepting,minimum size=1.2cm]  (D) [right of=C] {$\mathit{end}$};
	
	  \path (A) edge             node {$\phi_1$} (B)
	            edge[loop above] node {$\neg (\phi_1 \vee \phi_2 \vee \phi_3)$} (A)
	        (B) edge             node {$\phi_2 :: \pi_S$} (C)
	            edge[loop above] node {$\neg (\phi_1 \vee \phi_2 \vee \phi_3)$} (B)
	        (C) edge             node {$\phi_3~\land~\pi_I$} (D)
	            edge[loop above] node {$\neg (\phi_1 \vee \phi_2 \vee \phi_3)$} (C);
	\end{tikzpicture}
	\vspace{-5pt}
	{\small
	\begin{align*}
	\phi_1 \triangleq \ & \texttt{$G\langle$response\_type:code, redirect\_uri:\textasciicircum$\underbrace{\texttt{(}\overbrace{\text{\texttt{(https?://.*?/)}}}^{\text{\texttt{origin}}}\texttt{.*?)}}_{\text{\texttt{uri1}}}$(?:\textbackslash?|\$)$\rangle$} \\
	\phi_2 \triangleq \ & \texttt{$G($Location:[?\&]code=$\underbrace{\texttt{(.*?)}}_{\ident{authcode}}$(?:\&|\$)$)$} \qquad \phi_3 \triangleq \texttt{$\underbrace{\texttt{(.*)}}_{\texttt{uri2}}\langle$code:$\texttt{([\textasciicircum\textbackslash s]\{40,\})}\rangle$} \\
	\pi_S \triangleq \ & \ident{authcode} \rightarrow \{\texttt{https://accounts.google.com}, \ident{origin}\} \qquad \pi_I \triangleq \ \usage{uri1} = \usage{uri2}
	\end{align*}}
	\vspace{-20pt}
	\caption{Automaton for OAuth 2.0 (authorization code mode) where $G$ is the OAuth endpoint at Google.}
	\label{fig:oauth-dfa}
\end{figure*}
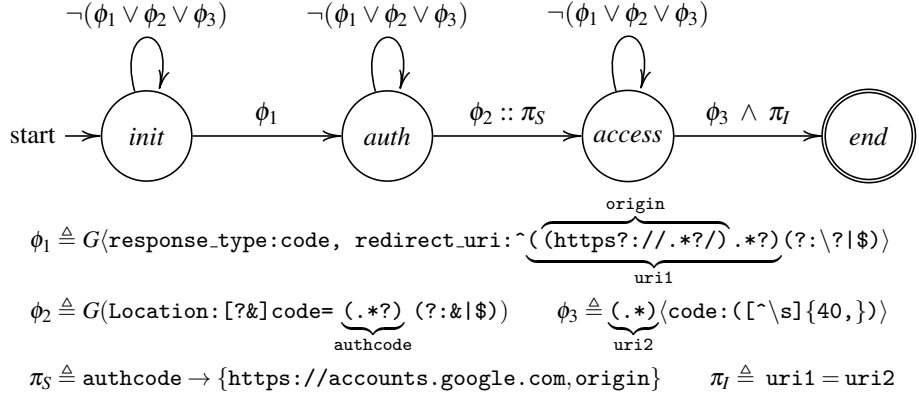

\WPE{} describes web protocols in terms of the HTTP(S) exchanges observed by the web browser, following the so-called \emph{browser relayed messages} methodology first introduced by Wang \etal~\cite{WangCW12}. The specification of the protocol flow defines the syntactic structure and the expected (sequential) order of the HTTP(S) messages, supporting the choice of different execution branches when a particular protocol message is sent or received by the browser. The protocol specification is given in XML (\cf{} Appendix~\ref{sec:xml}), but for the sake of readability, we use in this paper an equivalent representation in terms  of finite state automata, like the one depicted in Figure~\ref{fig:oauth-dfa}. Intuitively, each state of the automaton represents one stage of the protocol execution in the browser. By sending an HTTP(S) request or receiving an HTTP(S) response as dictated by the protocol, the automaton steps to the next state until it reaches a final state denoting the end of the protocol run. Afterwards, the automaton moves back to the initial state and a new protocol run can start.

The edges of the automaton are labeled with \emph{message patterns}, describing the expected shape of the protocol messages at each state. We represent HTTP(S) requests as $\reqshape{e}{a},$ where $e$ is the remote endpoint to which the message is sent and $a$ is a list of parameters, while HTTP(S) responses are noted $\respshape{e}{h}$, where $e$ is the remote endpoint from which the message is received and $h$ is a list of headers.\footnote{~We support HTTP headers also in requests. Here we omit them since they are not used in the protocols that we consider.} The syntactic structure of $e,a,h$ can be described using regular expressions. The message patterns should be considered as \emph{guards} of the transition, which are only enabled for messages matching the pattern. For instance, the pattern $\phi_2$ in Figure~\ref{fig:oauth-dfa} matches a response from the endpoint $G$ with a \texttt{Location} header that contains a URL with a parameter named \texttt{code}. If an HTTP(S) request or response does not satisfy any of the patterns of the outgoing transitions of the current state, it is blocked and the automaton is reset to the initial state, \ie, the protocol run is aborted. In case of branches with more than one transition enabled at a given state, we solve the non-determinism by picking the first transition (with a matching pattern) according to the order defined in the XML specification. Patterns can be composed using standard logical connectives.

Each state of the automaton also allows for pausing the protocol execution in presence of requests and responses that are unrelated to the protocol. Messages are considered unrelated to the protocol if they are not of the shape of any valid message in the protocol specification. In the automaton, this is expressed by having a self-loop for each state, labeled with the negated disjunction of all patterns describing valid protocol messages. This is important for website functionality, because the input/output behavior of browsers on realistic websites is complex and hard to fully determine when writing a protocol specification. Also, the same protocol may be run on different websites, which need to fetch different resources as part of their protocol-unrelated functionalities, and we would like to ensure that the same protocol specification can be enforced uniformly on all these websites.

\subsubsection{Security Policies}
To incorporate secrecy and integrity policies in the automaton, we allow for binding parts of message patterns to \emph{identifiers}. For instance, in Figure~\ref{fig:oauth-dfa} we bind the identifier \ident{origin} to the content of the \texttt{redirect\_uri} parameter, more precisely to the part matching the regular expression group \texttt{(https?://.*?/)}.\footnote{\hspace{3pt}\url{https://developer.mozilla.org/en-US/docs/Web/JavaScript/Reference/Global_Objects/RegExp}} The scope of an identifier includes the state where it is first introduced and all its successor states, where the notion of successor is induced by the tree structure of the automaton. For instance, the scope of the identifier \ident{origin} introduced in $\phi_1$ includes the states $\mathit{auth}, \mathit{access},\mathit{end}$.

The \emph{secrecy policy} defines which parts of the HTTP(S) responses included in the protocol specification must be confidential among a set of web origins. We express secrecy policies $\pi_S$ with the notation $\ident{x} \rightarrow S$ to denote that the value bound to the identifier \ident{x} can be disclosed only to the origins specified in the set $S$.  We call $S$ the \emph{secrecy set} of identifier \ident{x} and represent such a policy on the message pattern where the identifier \ident{x} is first introduced, using a double colon symbol $::$ as a separator. For instance, in Figure~\ref{fig:oauth-dfa} we require that the value of the authorization code, which is bound to the identifier \ident{authcode} introduced in $\phi_2$, can be disclosed only to Google (at \texttt{https://accounts.google.com}) and the relying party (bound to the identifier \ident{origin}). Confidential message components are stripped from HTTP(S) responses and substituted by random placeholders, so that they are isolated from browser accesses, \eg, computations performed by JavaScript. When the automaton detects an HTTP(S) request including one of the generated placeholders, it replaces the latter with the corresponding original value, but only if the HTTP(S) request is directed to one of the origins which is entitled to learn it. A similar idea was explored by Stock and Johns to strengthen the security of password managers~\cite{StockJ14}. Since the substitution of confidential message components with placeholders changes the content of the messages, potentially introducing deviations with respect to the transition labels, the automaton processes HTTP(S) responses before stripping confidential values and HTTP(S) requests after replacing the placeholders with the original values. This way, the input/output behavior of the automaton matches the protocol specification.

The \emph{integrity policy} defines runtime checks over the HTTP(S) messages. These checks allow for the comparison of incoming messages with the messages received in previous steps of the protocol execution. If any of the integrity checks fails, the corresponding message is not processed and the protocol run is aborted. To express integrity policies $\pi_I$ in the automaton, we enrich the message patterns to include comparisons ranging over the identifiers introduced by preceding messages. In the case of OAuth 2.0, we would like to ensure that the browser is redirected by the $\idp$ to the redirect URI specified in the first step of the protocol. Therefore, in Figure~\ref{fig:oauth-dfa} the desired integrity policy is modeled by the condition $\ident{uri1} = \ident{uri2}$.

\vspace{-5pt}

\subsubsection{Enforcing Multiple Protocols}
There are a couple of delicate points to address when multiple protocol specifications $P_1,\ldots,P_n$ must be enforced by \WPE:
\begin{enumerate}
\item if two different protocols $P_i$ and $P_j$ share messages with the same structure, there might be situations where \WPE does not know which of the two protocols is being run, yet a message may be allowed by $P_i$ and disallowed by $P_j$ or vice-versa;
\item if \WPE is enforcing a protocol $P_i$, it must block any message which may be part of another protocol $P_j$, otherwise it would be trivial to sidestep the security policy of $P_i$ by first making the browser process the first message of $P_j$.
\end{enumerate}
Both problems are solved by replacing the protocol specifications $P_1,\ldots,P_n$ with a single specification $P$ with $n$ branches, one for each $P_i$. Using this construction, any ambiguity on which protocol specification should be enforced is solved by the determinism of the resulting finite state automaton. Moreover, the self loops of the automaton will only match the messages which are not part of any of the $n$ protocol specifications, thereby preventing unintended protocol interleavings. Notice that the semantics of \WPE depends on the order of $P_1,\ldots,P_n$, due to the way we enforce determinism on the compiled automaton: if $P_i$ starts with a request to $u$ including two parameters $a$ and $b$, while $P_j$ starts with a request to $u$ including just the parameter $a$, then $P_i$ should occur before $P_j$ to ensure it is actually taken into account.

\subsection{Discussion}
\label{sec:limitations}
A number of points of the design and the implementation of \WPE{} are worth discussing more in detail.

\vspace{-4pt}

\subsubsection{Protocol Flow}
\WPE{} provides a significant improvement in security over standard web browsers, as we show in the remainder of the paper, but the protection it offers is not for free, because it requires the specification of a protocol flow and a security policy. We think that it is possible to develop automated techniques to reconstruct the intended protocol flow from observable browser behaviours, while synthesizing the security policy looks more difficult. Manually finding the best security policy for a protocol may require significant expertise, but even simple policies can be useful to prevent a number of dangerous attacks, as we demonstrate in Section~\ref{sec:protocols}.

The specification style of the protocol flow supported by \WPE{} is simple, because it only allows sequential composition of messages and branching. As a result, our finite state automata are significantly simpler than the request graphs proposed by Guha \etal~\cite{GuhaKJ09} to represent legitimate browser behaviors (from the server perspective). For instance, our finite state automata do not include loops and interleaving of messages, because it seems that these features are not extensively used in web protocols. Like standard security protocols, web protocols are typically specified in terms of a fixed number of sequential messages, which are appropriately supported by the specification language we chose. 

\subsubsection{Secrecy Enforcement}
The implementation of the secrecy policies of \WPE{} is robust, but restrictive. Since \WPE{} substitutes confidential values with random placeholders, only the latter are exposed to browser-side scripts. Shielding secret values from script accesses is crucial to prevent confidentiality breaches via untrusted scripts or XSS, but it might also break the website functionality if a trusted script needs to compute over a secret value exchanged in the protocol. The current design of \WPE{} only supports a limited use of secrets by browser-side scripts, \ie, scripts can only forward secrets unchanged to the web origins entitled to learn them. We empirically show that this is enough to support existing protocols like OAuth 2.0 and SAML, but other protocols may require more flexibility.

Dynamic information flow control deals with the problem of letting programs compute over secret values while avoiding confidentiality breaches and it has been applied in the context of web browsers~\cite{GroefDNP12,HedinBS15,BichhawatRGH14,RajaniB0015,BauerCJPST15}. We believe that dynamic information flow control can be fruitfully combined with \WPE{} to support more flexible secrecy policies. This integration can also be useful to provide confidentiality guarantees for values which are generated at the browser-side and sent in HTTP(S) requests, rather than received in HTTP(S) responses. We leave the study of the integration of dynamic information flow control into \WPE{} to future work.

\vspace{-5pt}

\subsubsection{Extension APIs}
The current prototype of \WPE{} suffers from some limitations due to the Google Chrome extension APIs. In particular, the body of HTTP messages cannot be modified by extensions, hence the secrecy policy cannot be implemented when secret values are embedded in the page contents or the corresponding placeholders are sent as POST parameters. Currently, we protect secret values contained in the HTTP headers of a response (\eg, cookies or parameters in the URL of a \texttt{Location} header) and we only substitute the corresponding placeholders when they are communicated via HTTP headers or as URL parameters. Clearly this is not a limitation of our general approach but rather one of the extension APIs, which can be solved by implementing the security monitor directly in the browser or as a separate proxy application. Despite these limitations, we were able to test the current prototype of \WPE{} on a number of real-world websites with very promising results, as reported in Section~\ref{sec:experiments}.

\section{Fortifying Web Protocols with \WPE}
\label{sec:protocols}
% !TeX root = main.tex

To better appreciate the security guarantees offered by \WPE, we consider two popular web protocols: OAuth 2.0 and SAML. The security of both protocols has already been studied in depth, so they are an excellent benchmark to assess the effectiveness of \WPE: we refer to~\cite{BansalBDM14,FettKS16,SunB12} for security analyses of OAuth 2.0 and to~\cite{ArmandoCCCPS13,ArmandoCCCT08} for research studies on SAML. Remarkably, by writing down a precise security policy for SAML, we were able to expose a new critical attack against the Google implementation of the protocol.

\subsection{Attacks Against OAuth 2.0}
\label{subsec:oauth}

\begin{table}[t]
	\centering
	\begin{tabular}{cp{5.75cm}}
		\multicolumn{1}{p{1.4cm}}{\parbox{1.4cm}{\centering\textbf{Detected Violation}}} & \multicolumn{1}{c}{\textbf{Attack}} \\[7pt]
		\toprule
		\multicolumn{1}{c}{\multirow{3}{*}{\parbox{1.4cm}{\centering Protocol flow \\ deviation}}}	 & Session swapping~\cite{SunB12} \\
		& Social login CSRF on stateless clients~\cite{BansalBDM14} \\
		& IdP mix-up attack (web attacker)~\cite{FettKS16} \\
		\midrule
		\multicolumn{1}{c}{\multirow{6}{*}{\parbox{1.4cm}{\centering Secrecy \\ violation}}} & Unauthorized login by authentication code redirection~\cite{BansalBDM14} \\
		& Resource theft by access token redirection~\cite{BansalBDM14} \\
		& 307 redirect attack~\cite{FettKS16} \\
		& State leak attack~\cite{FettKS16} \\
		\midrule
		\multicolumn{1}{c}{\multirow{2}{*}{\parbox{1.4cm}{\centering Integrity \\ violation}}} & Cross social-network request forgery~\cite{BansalBDM14} \\
		& Na\"ive RP session integrity attack~\cite{FettKS16} \\
		\bottomrule
	\end{tabular}
	\caption{Overview of the attacks against OAuth 2.0.}
	\label{tab:attacks}
\end{table}
		
We review in this section several attacks on OAuth 2.0 from the literature, analysing whether they are prevented by our extension. We focus in particular on those presented in~\cite{BansalBDM14,FettKS16,SunB12}, since they apply to the OAuth 2.0 flows presented in this work. In Table~\ref{tab:attacks} we provide an overview of the attacks that \WPE{} is able to prevent, grouped according to the type of violation of the security properties that they expose. 

\subsubsection{Protocol Flow Deviations} 
This category covers attacks that force the user's browser to skip messages or to accept them in a wrong order.  For instance, some attacks, \eg, some variants of CSRF and session swapping, rely on completing a social login in the user's browser that was not initiated before. This is a clear deviation from the intended protocol flow and, as a consequence, \WPE{} blocks these attacks.

We exemplify on the session swapping attack discussed in Section~\ref{subsec:sessionswapping}. Here the attacker tricks the user into sending a request containing the attacker's authorization credential (\eg, the authorization code) to $\rp$ (step \circled{4} of the protocol flow). Since the state parameter is not used, the $\rp$ cannot verify whether this request was preceded by a social login request by the user. Our security monitor blocks the (out-of-order) request since it matches the pattern $\phi_3$, which is allowed by the automaton in Figure~\ref{fig:oauth-dfa} only in state \state{access}. Thus, the attack is successfully prevented.

\subsubsection{Secrecy Violations}
This category covers attacks where sensitive information is unintentionally leaked, \eg, via the Referer header or because of the presence of open redirectors at $\rp$. Sensitive data can either be leaked to untrusted third parties that should not be involved in the protocol flow (as in the state leak attack) or protocol parties that are not trusted for a specific secret (as in the 307 redirect attack). \WPE{} can prevent this class of attacks since the secrecy policy allows one to specify the origins that are entitled to receive a secret. 

We illustrate how the monitor prevents these attacks in case of the state leak attack discussed in Section~\ref{subsec:stateleak}, focusing on the authorization code. In the attack, the authorization code is leaked via the Referer header of the request fetching a resource from the attacker website which is embedded in the page located at the redirect URI of $\rp$ (step \circled{4} of the protocol).
When the authorization code (\ident{authcode}) is received (step \circled{2}), 
the monitor extracts it from the Location header and replaces it with a random placeholder before the request is processed by the browser. After step \circled{4}, the request to the attacker's website is sent, but the monitor does not replace the placeholder with the actual value of the authorization code since the secrecy set associated to \ident{authcode} in $\pi_S$ does not include the domain of the attacker.

\subsubsection{Integrity Violations} 
This category contains attacks that maintain the general protocol flow, but the contents of the exchanged messages do not satisfy some integrity constraints required by the protocol. \WPE{} can prevent these attacks by enforcing browser-side integrity checks.

Consider the na\"ive RP session integrity attack presented in Section~\ref{subsec:naivesessionintegrity}. In this attack, the malicious identity provider $\aidp$ redirects the user's browser to the redirect URI of the honest identity provider $\hidp$ at $\rp$ during step \circled{4} of the protocol. At step \circled{2}, the redirect URI is provided to $\aidp$ as parameter. This request corresponds to the pattern $\phi_1$ of the automation and the redirect URI associated to $\aidp$ is bound to the identifier \ident{uri1}. At step \circled{4}, $\aidp$ redirects the browser to a different redirect URI, which is bound to the identifier \ident{uri2}. Although the shape of the request satisfies pattern $\phi_3$, the monitor cannot move from state \state{access} to state \state{end} since the constraint \texttt{\ident{uri1} = \ident{uri2}} in the integrity policy $\pi_I$ is violated. Thus, no transition is enabled for the state \state{access} and the request is blocked by \WPE, therefore preventing the attack.

\subsection{Attacks Against SAML}

\tikzset{>={To[length=2.5mm,width=1.5mm]}}
\begin{figure*}[tb]
	\centering
	\def\hd{0.6cm}
	\begin{tikzpicture}[node distance=7cm and 1cm,  font=\ttfamily\footnotesize]
		\node(user){$\webbrowser$};
		\node[right of = user](rp){$\sep$}; 
		\node[right of = rp](idp){$\idp$}; 
		
		\path[draw, ->] ($(user)-(0, \hd)$) to node[above]{\circled{1} URI} ($(rp)-(0, \hd)$); 
		\path[draw, -] ($(user)-(0, 2*\hd)$) to node[above]{} ($(rp)-(0, 2*\hd)$); 
		\path[draw, -] ($(user)-(0, 2*\hd)$) to ($(user)-(0, 3*\hd)$); 
		\path[draw, ->] ($(user)-(0, 3*\hd)$) to node[above]{\circled{2} SAMLRequest=AuthnRequest, RelayState=URI} ($(idp)-(0, 3*\hd)$); 
		\path[draw, <-] ($(user)-(0, 4*\hd)$) to node[above]{\circled{3} login form} ($(idp)-(0, 4*\hd)$); 
		\path[draw, ->] ($(user)-(0, 5*\hd)$) to node[above]{User credentials} ($(idp)-(0, 5*\hd)$); 
		\path[draw, -] ($(user)-(0, 6*\hd)$) to node[above]{} ($(idp)-(0, 6*\hd)$); 
		\path[draw, -] ($(user)-(0, 6*\hd)$) to ($(user)-(0, 7*\hd)$); 
		\path[draw, ->] ($(user)-(0, 7*\hd)$) to node[above]{\circled{4} SAMLResponse=Response, RelayState=URI} ($(rp)-(0, 7*\hd)$); 

		\path[draw, -] ($(user)-(0, 8*\hd)$) to node[above]{} ($(rp)-(0, 8*\hd)$); 
		\path[draw, -] ($(user)-(0, 8*\hd)$) to ($(user)-(0, 9*\hd)$); 
		\path[draw, ->] ($(user)-(0, 9*\hd)$) to node[above]{\circled{5} URI} ($(rp)-(0, 9*\hd)$); 

		\path[draw, <-] ($(user)-(0, 10*\hd)$) to node[above]{\circled{6} resource} ($(rp)-(0, 10*\hd)$); 
	\end{tikzpicture}
	\caption{SAML 2.0 SP-Initiated SSO with Redirect/POST Bindings.}
	\label{fig:saml-sso}
\end{figure*}
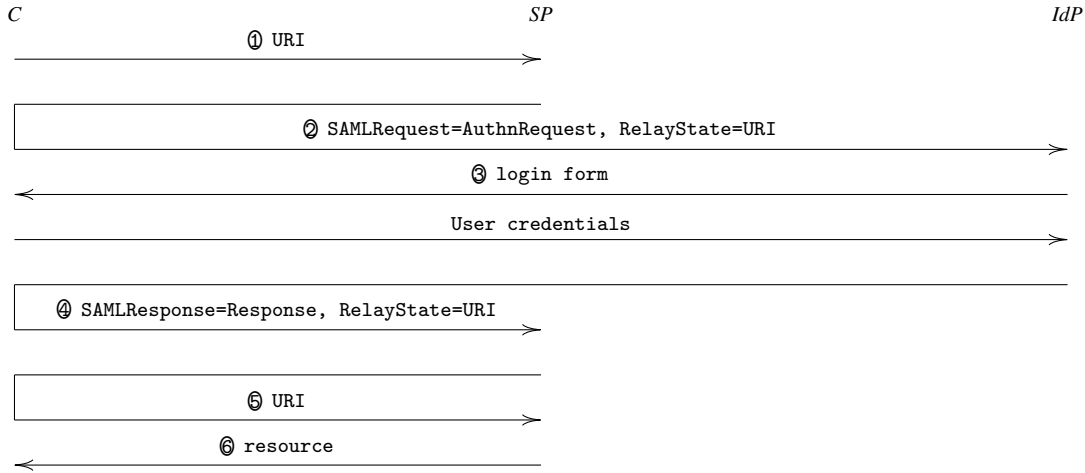

The \emph{Security Assertion Markup Language} (SAML) 2.0~\cite{SAML} is an open standard for sharing authentication and authorization across a multitude of domains. SAML is based on XML messages called \emph{assertions} and defines different \emph{profiles} to account for a variety of use cases and deployment scenarios. SSO functionality is enabled by the SAML 2.0 web browser SSO profile, whose typical use case is the $\sep$-Initiated SSO with Redirect/POST Bindings~\cite{SAMLprofiles,ArmandoCCCT08}. Similarly to OAuth 2.0, there are three entities involved: a user controlling a web browser ($\webbrowser$), an identity provider ($\idp$) and a service provider ($\sep$). The protocol prescribes how $\webbrowser$ can access a resource provided by an $\sep$ after authenticating with an $\idp$.

The relevant steps of the protocol are depicted in Figure~\ref{fig:saml-sso}. In step \circled{1}, $\webbrowser$ requests from $\sep$ the resource located at $\params{URI}$; in \circled{2} the $\sep$ redirects the browser to the $\idp$ sending an \emph{AuthnRequest} XML message in deflated, base64-encoded form and a \emph{RelayState} parameter; $\webbrowser$ provides his credentials to the $\idp$ in step \circled{3} where they are verified; in step \circled{4} the $\idp$ causes the browser to issue a POST request to the Assertion Consumer Service at the $\sep$ containing the base64-encoded \emph{SamlResponse} and the \emph{RelayState} parameters; in \circled{5} the $\sep$ processes the response, creates a security context at the service provider and redirects $\webbrowser$ to the target resource at $\params{URI}$; given that a security context is in place, the $\sep$ provider returns the resource to $\webbrowser$.

The RelayState is a mechanism for preserving some state information at the $\sep$, such as the resource URI requested by the user~\cite{GoogleSAML}. If the RelayState parameter is used within a request message, then subsequent responses must maintain the exact value received with the request~\cite{SAMLbindings}. A violation of this constraint enables attacks such as~\cite{ArmandoCCCPS13}, in which $\webbrowser$ requests a resource $\params{URI}_i$ at a malicious $\sep_i$. $\sep_i$ pretends to be $\webbrowser$ at the honest $\sep$ and requests a different resource at $\sep$ located at $\params{URI}_\sep$ which is returned to $\sep_i$. The malicious service provider replies to $\webbrowser$ by providing a redirection address containing a different resource $\params{URI}$, thus causing the browser to send $\params{URI}_i$ instead of instead of $\params{URI}$ as the value of RelayState at steps \circled{2},\circled{4}. The result is that $\webbrowser$ forcibly accesses a resource at $\sep$, while he originally asked for a resource from $\sep_i$.

Interestingly, by using WPSE it is possible to instruct the browser with knowledge of the protocol in such a way that the client can verify whether the requests at steps \circled{2},\circled{4} are related to the initial request.  We distilled a simple policy for the SAML 2.0 web browser SSO profile that enforces an integrity constraint on the value of the RelayState parameter, thus blocking requests to undesired resources due to a violation of the policy.

Furthermore, SAML 2.0 does not specify any way to maintain a contextual binding between the request at step \circled{2} and the request at step \circled{4}. It follows that only the \emph{SAMLResponse} and \emph{RelayState} parameters are enough to allow $\webbrowser$ to access the resource at $\params{URI}$. We discovered that this shortcoming in the protocol has a critical impact on real $\sep$s using the SAML-based SSO profile described in this section. Indeed, we managed to mount an attack against Google that allows a web attacker to authenticate any user on Google's suite applications under the attacker's account, with effects similar to a Login CSRF attack. Since Google can act as a Service Provider ($\sep$) with a third party $\idp$, an attacker registered to a given $\idp$ can simulate a login attempt with his legitimate credentials to obtain a valid POST request to the Google assertion consumer service (step \circled{4}). Once accessed, a malicious web page can then cause a victim's browser to issue the attacker's request to the Google assertion consumer service, thus forcing the victim inside the attacker's controlled authenticated session.

The vulnerability can be exploited by any web attacker with a valid account on a third party $\idp$ that uses Google as $\sep$. In particular, our university uses SAML 2.0 with Google as a Service provider to offer email and storage facilities to students and employees. We have implemented the attack by constructing a malicious webpage that silently performs a login on Google's suite applications using one of our personal accounts. The vulnerability allows the attacker to access private information of the victim that has been saved in the account, such as activity history, notes and documents. We have responsibly reported this vulnerability to Google who rewarded us according to their bug bounty program. As soon as they are available, we will provide on our website the details of the fixes that Google is implementing to resolve the issue~\cite{BlogPostSAML}.

From the browser standpoint, this attack is clearly caused by a violation of the protocol flow given that steps \circled{1}-\circled{3} are carried out by the attacker and step \circled{4} and subsequent ones involve the victim. \WPE identifies the outgoing request to the $\idp$ as a protocol flow deviation, thereby preventing the attack. 

\subsection{Out-of-Scope Attacks}
We have shown that \WPE{} is able to block a wide range of attacks on existing web protocols. However, some classes of attacks cannot be prevented by browser-side security monitoring. Specifically, \WPE{} cannot prevent:
\begin{enumerate}
\item attacks which do not deviate from the expected protocol flow. An example of such an attack against OAuth 2.0 is the \emph{automatic login CSRF} attack presented in~\cite{BansalBDM14}, which exploits the lack of CSRF protection on the login form of the relying party to force an authentication to the identity provider. This class of attacks can be prevented by implementing appropriate defenses against known web attacks;
\item attacks which cause deviations from the expected protocol flow that are not observable by the browser. In particular, this class of attacks includes \emph{network attacks}, where the attacker corrupts the traffic exchanged between the protocol participants. For instance, a network attacker can run the \emph{IdP mix-up} attack from~\cite{FettKS16} when the first step of OAuth 2.0 is performed over HTTP. This class of attacks can be prevented by making use of HTTPS, preferably backed up by HSTS;
\item attacks which do not involve the user's browser at all. An example is the \emph{impersonation} attack on OAuth 2.0 discussed in~\cite{SunB12}, where public information is used for authentication. Another example is the \emph{DuoSec} vulnerability found on several SAML implementations~\cite{DuosecSAML} that exploits a bug in the XML libraries used by SPs to parse SAML messages. This class of attacks must be necessarily solved at the server side.
\end{enumerate}

\section{Experimental Evaluation}
\label{sec:experiments}
% !TeX root = main.tex
Having discussed how \WPE{} can prevent several real-world attacks presented in the literature, we finally move to on-field experiments. The goal of the present section is assessing the practical security benefits offered by \WPE{} on existing websites in the wild, as well as to test the compatibility of its browser-side security monitoring with current web technologies and programming practices. To this end, we experimentally assessed the effectiveness of \WPE{} by testing it against websites using OAuth 2.0 to implement SSO at high-profile $\idp$s.

\subsection{Experimental Setup}
We developed a crawler to automatically identify existing OAuth 2.0 implementations in the wild. Our analysis is not meant to provide a comprehensive coverage of the deployment of OAuth 2.0 on the web, but just to identify a few popular identity providers and their relying parties to carry out a first experimental evaluation of \WPE. 

We started from a comprehensive list of OAuth 2.0 identity providers\footnote{\hspace{3pt}\url{https://en.wikipedia.org/wiki/List_of_OAuth_providers}} and we collected for each of them the list of the HTTP(S) endpoints used in their implementation of the protocol. Inspired by~\cite{AckerHS17}, our crawler looks for login pages on websites to find syntactic occurrences of these endpoints: after accessing a homepage, the crawler extracts a list of (at most) 10 links which may likely point to a login page, using a simple heuristic. It also retrieves, using the Bing search engine, the 5 most popular pages of the website. For all these pages, the crawler checks for the presence of the OAuth 2.0 endpoints in the HTML code and in the 5 topmost scripts included by them. By running our crawler on the Alexa 100k top websites, we found that Facebook (1,666 websites), Google (1,071 websites) and VK (403 websites) are the most popular identity providers in the wild.

We then developed a faithful XML representation of the OAuth 2.0 implementations available at the selected identity providers. There is obviously a large overlap between these specifications, though slight differences are present in practice, \eg, the use of the \texttt{response\_type} parameter is mandatory at Google, but can be omitted at Facebook and VK to default to the authorization code mode. For the sake of simplicity, we decided to model the most common use case of OAuth 2.0, \ie, we assume that the user has an ongoing session with the identity provider and that authorization to access the user's resources on the provider has been previously granted to the relying party. For each identity provider we devised a specification that supports the OAuth 2.0 authorization code and implicit modes, with and without the optional state parameter, leading to 4 possible execution paths. Finally, we created a dataset of 90 websites by sampling 30 relying parties for each identity provider, covering both the authorization code mode and the implicit mode of OAuth 2.0. We have manually visited these websites with a browser running \WPE{} both to verify if the protocol run was completed successfully and to assess whether all the functionalities of the sites were working properly. In the following we report on the results of testing our extension against these websites from both a security and a compatibility point of view.

\subsection{Security Analysis}
We devised an automated technique to check whether \WPE{} can stop dangerous real-world attacks. Since we did not want to attack the websites, we focused on two classes of vulnerabilities which are easy to detect just by navigating the websites when using \WPE. The first class of vulnerabilities enables confidentiality violations: it is found when one of the placeholders generated by \WPE{} to enforce its secrecy policies is sent to an unintended web origin. The second class of vulnerabilities, instead, is related to the use of the state parameter: if the state parameter is unused or set to a predictable static value, then session swapping becomes possible (see Section~\ref{subsec:sessionswapping}). We can detect these cases by checking which protocol specification is enforced by \WPE{} and by making the state parameter secret, so that all the values bound to it are collected by \WPE{} when they are substituted by the placeholders used to enforce the secrecy policy.

We observed that our extension prevented the leakage of sensitive data on 4 different relying parties. Interestingly, we found that the security violation exposed by the tool are in all cases due to the presence of tracking or advertisements libraries such as Facebook Pixel,\footnote{\hspace{3pt}\url{https://www.facebook.com/business/a/facebook-pixel}} Google AdSense,\footnote{~\url{https://www.google.com/adsense}} Heap\footnote{~\url{https://heapanalytics.com/}} and others. For example, this has been observed on \url{ticktick.com}, a website offering collaborative task management tools. The leakage is enabled by two conditions:
\begin{enumerate}
\item the website allows its users to perform a login via Google using the implicit mode;  
\item the Facebook tracking library is embedded in the page which serves as redirect URI.
\end{enumerate}
Under these settings, right after step \circled{4} of the protocol, the tracking library sends a request to \url{https://www.facebook.com/tr/} with the full URL of the current page, which includes the access token issued by Google. We argue that this is a critical vulnerability, given that leaking the access token to an unauthorized party allows unintended access to sensitive data owned by the users of the affected website. We promptly reported the issue to the major tracking library vendors and the vulnerable websites. Library vendors informed us that they are not providing any fix since it is a responsibility of web developers to include the tracking library only in pages without sensitive contents.\footnote{~See, for instance, Google AdSense program policy available at \url{https://support.google.com/adsense/topic/6162392}}

For what concerns the second class of vulnerabilities, 55 out of 90 websites have been found affected by the lack or misuse of the state parameter. More in detail, we identified 41 websites that do not support it, while the remaining 14 websites miss the security benefit of the state parameter by using a predictable or constant string as a value. We claim that such disheartening situation is mainly caused by the identity providers not setting this important parameter as mandatory. In fact, the state parameter is listed as recommended by Google and optional by VK. On the other hand, Facebook marks the state parameter as mandatory in its documentation, but our experiments showed that it fails to fulfill the requirement in practice. Additionally, it would be advisable to clearly point out in the OAuth 2.0 documentation of each provider the security implications of the parameter. For instance, according to the Google documentation,\footnote{~\url{https://developers.google.com/identity/protocols/OAuth2WebServer}} the state parameter can be used ``for several purposes, such as directing the user to the correct resource in your application, sending nonces, and mitigating cross-site request forgery'': we believe that this description is too vague and opens the door to misunderstandings.

\subsection{Compatibility Analysis}
To detect whether \WPE{} negatively affects the web browser functionality, we performed a basic navigation session on the websites in our dataset. This interaction includes an access to their homepage, the identification of the SSO page, the execution of the OAuth 2.0 protocol, and a brief navigation of the private area of the website. In our experiments, the usage of \WPE{} did not impact in a perceivable way the browser performance or the time required to load webpages. We were able to navigate 81 websites flawlessly, but we also found 9 websites where we did not manage to successfully complete the protocol run.

In all the cases, the reason for the compatibility issues was the same, \ie, the presence of an HTTP(S) request with a parameter called \texttt{code} after the execution of the protocol run. This message has the same syntactic structure as the last request sent as part of the authorization code mode of OAuth 2.0 and is detected as an attack when our security monitor moves back to its initial state at the end of the protocol run, because the message is indistinguishable from a session swapping attempt (see Section~\ref{subsec:sessionswapping}). We manually investigated all these cases: 2 of them were related to the use of the Gigya social login provider, which offers a unified access interface to many identity providers including Facebook and Google; the other 7, instead, were due to a second exchange of the authorization code at the end of the protocol run. We were able to solve the first issue by writing an XML specification for Gigya (limited to Facebook and Google), while the other cases openly deviate from the OAuth 2.0 specification, where the authorization code is only supposed to be sent to the redirect URI and delivered to the relying party from there. These custom practices are hard to explain and to support and, unsurprisingly, may introduce security flaws. In fact, one of the websites deviating from the OAuth 2.0 specification suffers from a serious security issue, because the authorization code is first communicated to the website over HTTP before being sent over HTTPS, thus becoming exposed to network attackers. We responsibly disclosed this security issue to the website owners.

In the end, all the compatibility issues we found boil down to the fact that a web protocol message has a relatively weak syntactic structure, which may end up matching a custom message used by websites as part of their functionality. We think that most of these issues can be robustly solved by using more explicit message formats for standardized web protocols like OAuth 2.0: explicitness is indeed a widely recognized prudent engineering practice for traditional security protocols~\cite{AbadiN96}. Having structured message formats could be extremely helpful for a precise browser-side fortification of web protocols which minimizes compatibility issues.

\section{Formal Guarantees}
\begin{figure}[t]
\iffull
\centering
\fi
\includegraphics[scale=0.24]{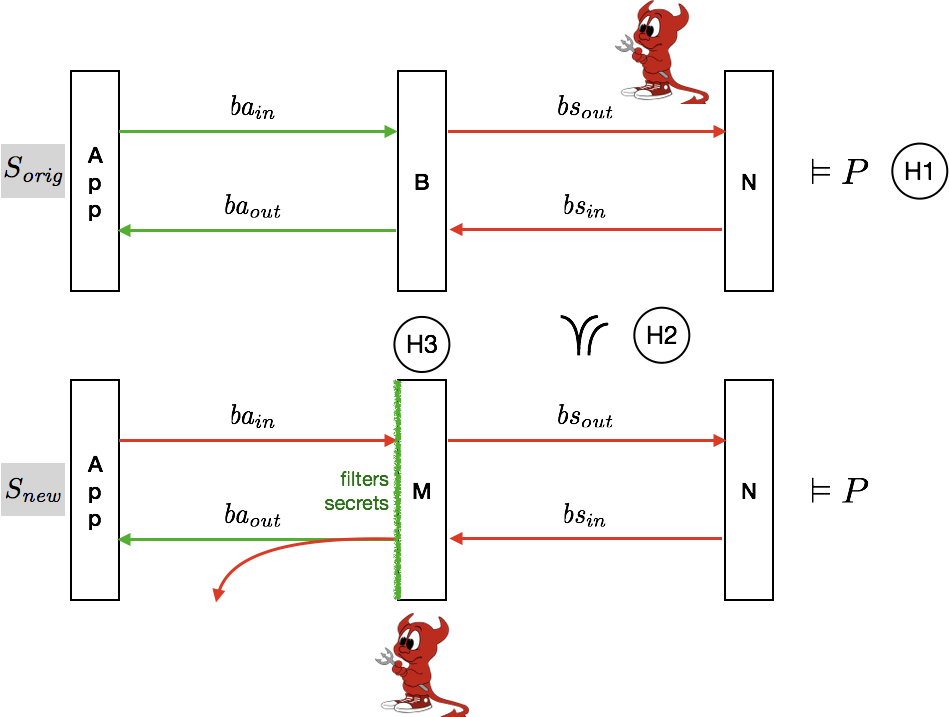}
\caption{Visual description of Theorem~\ref{thm:correctness}}
\label{fig:theorem}
\end{figure}
Now we formally characterize the security guarantees offered by our monitoring technique.  
\iffull
First we provide an intuitive description of the  formal result followed by a detailed description of the formalization. 
\else
Here we provide an intuitive description of the  formal result, referring the interested reader to~\cite{techreport} for a complete account. 
\fi

The formal result states that given a web protocol that is proven secure for a set of network participants and an uncorrupted client, by our monitoring approach we can achieve the same security guarantees given a corrupted client (\eg, due to XSS attacks). More precisely this means that all attacks that will not occur in the presence of an ideally behaving client can be fixed by our monitor. 
Of course, these security guarantees only span the run of the protocol that is proven secure and its protocol-specific secrets. So the monitor can \eg, ensure that the OAuth 2.0 protocol is securely executed in the presence of compromised scripts which might result in successful authentication and the setting of a session cookie. However, the monitor cannot prevent that this session cookie is leaked by a malicious script after the protocol run is over. So other security techniques (\eg, the \texttt{HttpOnly} attribute for cookies) have to be in place or the protocol specification can in principle be extended to include the subsequent application steps (\eg, we can protect session cookies like we do for access tokens).

Our theory is elaborated within the applied pi calculus~\cite{RyanS11}, a popular process calculus for the formal analysis of cryptographic protocols, which is supported by various automated cryptographic protocol verifiers, such as ProVerif~\cite{Blanchet01}. Bansal \etal~\cite{BansalBDM14} have recently presented a technique to leverage ProVerif for the analysis of web protocol specifications, including OAuth. 

We give an overview on the theorem in Figure~\ref{fig:theorem}.  We assume that the protocol specification has already been proven secure in a setting where the browser-side application is well-behaved and, in particular, follows the protocol specification ($\orig$). Intuitively, our theorem says that security carries over to a setting ($\monitored$) where the browser-side application is totally under the control of the attacker (\eg, because of XSS attacks or a simple bug in the code) but the communication between the browser and the other protocol parties is mediated by our monitor. 

Specifically, $\orig$ includes  a  browser $\browser$ and an uncompromised application $\app$, which exchange messages via  private (green) communication channels $\bappin, \bappout$. The communication between the browser $\browser$ and the network $\network$ is performed via the public  (red) channels $\bservin, \bservout$ that can be observed and infiltrated by the network attacker.
$\monitored$ shows the setting in which the application is compromised: channel $\bappin$ for requests from the application to the browser is made public, modeling that arbitrary requests can be performed on it by the attacker. In addition, we assume the channel $\bappout$ modeling the responses from the browser to the app to leak all messages and consequently modeling that the compromised application might leak these secrets. Indeed, the compromised application can communicate with the network attacker, which can in turn use the learned information to attack the protocol. 

We state a simplified version of the correctness theorem as follows: 
\begin{theorem}[Monitor Correctness]
\label{thm:correctness}
Let processes $\app$, $\network$, $\browser$ and $\monitor$ as defined in $\orig$ and $P$ be a property on execution traces against a network attacker. 
Assume that the following conditions hold: 
\begin{enumerate}[label=\textsf{(H\arabic*)}]
\item \label{item:h1} $\orig \vDash P$ \quad  (`$\orig$ satisfies $P$') 
\item \label{item:h2} $\projected{\monitor}{\bservin, \bservout} ~\subsumes ~\projected{\orig}{\bservin, \bservout}$
\quad (`the set of requests/responses  on $\bservin$,$\bservout$ allowed by $\monitor$ are a subset of those produced by $\orig$')
\item \label{item:h3} $\monitor$ does not leak any secrets (\ie, messages initially unknown to the attacker) on $\bappout$
\end{enumerate}
Then it also holds that:
\begin{enumerate}[label=\textsf{(C)}]
\item \label{item:conc} $\monitored \vDash P$ \quad (`$\monitored$ satisfies $P$').
\end{enumerate}
\end{theorem}

Assumption~\ref{item:h1} states that the process as shown in $\orig$ satisfies a certain trace property. In the applied pi calculus, this is modeled by requiring that each partial execution trace of $\orig$ in parallel with an arbitrary network attacker satisfies the trace predicate $P$. 
Assumption~\ref{item:h2} states that the requests/responses allowed by the monitor $\monitor$ on the channels $\bservin$, $\bservout$, which model the communication between the browser and the network, are a subset of those possibly performed by the process $\orig$. Intuitively, this means that the monitor allows for the intended protocol flow, filtering out messages deviating from it.  Formally this is captured by projecting the execution traces of the corresponding processes to those components that model the input and output behavior on $\bservin$ and $\bservout$ and by requiring that for every such execution trace of $\monitor$ there is a corresponding one for $\orig$. 
Finally, assumption~\ref{item:h3} states that  the monitor $\monitor$ should not leak any secrets with its outputs on channel $\bappout$. In applied pi calculus this is captured by requiring that the outputs of $\monitor$ on channel $\bappout$ do not to contain any information that increases the attacker knowledge. 

Together these assumptions ensure that the monitored browser behaves as the ideal protocol participant in $\orig$ towards the network and additionally assure that an attacker cannot gain any additional knowledge via a compromised application that could enable her to perform attacks against the protocol over the network.
Formally, this is captured in conclusion~\ref{item:conc} that requires the partial execution traces of $\monitored$ to satisfy the trace predicate $P$. 

\subsection{Discussion}
Our formal result is interesting for various reasons. First, it allows us to establish formal security guarantees in a stronger attacker model by checking certain semantic conditions on the monitor, without having to prove from scratch the security of the protocol with the monitor in place on the browser-side. Second, the theorem demonstrates that enforcing the three security properties identified in Section~\ref{sec:challenges} does indeed suffice to protect web protocols from a large class of bugs and vulnerabilities on the browser side:~\ref{item:h2} captures the compliance with the intended protocol flow as well as data integrity, while~\ref{item:h3} characterizes  the secrecy of messages.

Finally, the three hypotheses of the theorem are usually extremely easy to check. For instance, let us consider the OAuth protocol. As previously mentioned, this has been formally analyzed in~\cite{BansalBDM14}, so~\ref{item:h1} holds true. In particular, the intended protocol flow is directly derivable from the applied pi calculus specification. The automaton in Figure~\ref{fig:oauth-dfa} only allows for the intended protocol flow, which is clearly contained in the execution traces analyzed in ~\cite{BansalBDM14}. Hence~\ref{item:h2} holds true as well. Finally, the only secrets in the protocol specification are those subject to the confidentiality policy in the automaton in Figure~\ref{fig:oauth-dfa}: as previously mentioned, these are replaced by placeholders, which are then passed to the web application. Hence no secret can ever leak, which validates~\ref{item:h3}.

\section{Related Work}
\subsection{Analysis of Web Protocols}
The first paper to highlight the differences between web protocols and traditional cryptographic protocols is due to Gross \etal~\cite{GrossPS05}. The paper presented a model of web browsers, based on a formalism reminiscent of input/output automata, and applied it to the analysis of password-based authentication, a key ingredient of most browser-based protocols. The model was later used to formally assess the security of the WSFPI protocol~\cite{GrobetaPS05}.

Traditional protocol verification tools have been successfully applied to find attacks in protocol specifications. For instance, Armando \etal{} analyzed both the SAML protocol and a variant of the protocol implemented by Google using the SATMC model-checker~\cite{ArmandoCCCT08}. Their analysis exposed an attack against the authentication goals of the Google implementation. Follow-up work by the same group used a more accurate model to find an authentication flaw also in the original SAML specification~\cite{ArmandoCCCPS13}. Akhawe \etal{} used the Alloy framework to develop a core model of the web infrastructure, geared towards attack finding~\cite{AkhaweBLMS10}. The paper studied the security of the WebAuth authentication protocol among other case studies, finding a login CSRF attack against it. The WebSpi library for ProVerif by Bansal \etal{} has been successfully applied to find attacks against existing web protocols, including OAuth 2.0~\cite{BansalBDM14} and cloud storage protocols~\cite{BansalBDM13}. Fett \etal{} developed the most comprehensive model of the web infrastructure available to date and fruitfully applied it to the analysis of a number of web protocols, including BrowserID~\cite{FettKS14}, SPRESSO~\cite{FettKS15} and OAuth 2.0~\cite{FettKS16}. 

Protocol analysis techniques are useful to verify the security of protocols, but they assume websites are correctly implemented and do not depart from the specification, hence many security researchers performed empirical security assessments of existing web protocol implementations, finding dangerous attacks in the wild. Protocols which deserved attention by the research community include SAML~\cite{SomorovskyMSKJ12}, OAuth 2.0~\cite{SunB12,LiM14} and OpenID Connect~\cite{LiM16}. Automated tools for finding vulnerabilities in web protocol implementations have also been proposed by security researchers~\cite{WangCW12,ZhouE14,YangLLZH16,MainkaMSW17}. None of these works, however, presented a technique to protect users accessing vulnerable websites in their browsers.

\subsection{Security Automata}
The use of finite state automata for security enforcement is certainly not new. The pioneering work in the area is due to Schneider~\cite{Schneider00}, which first introduced a formalization of security automata and studied their expressive power in terms of a class of enforceable policies. Security automata can only stop a program execution when a policy violation is detected; later work by Ligatti \etal{} extended the class of security automata to also include edit automata, which can suppress and insert individual program actions~\cite{LigattiBW05}. Edit automata have been applied to the web security setting by Yu \etal, who used them to express security policies for JavaScript code~\cite{YuCIS07}. The focus of their paper, however, is not on web protocols and is only limited to JavaScript, because input/output operations which are not JavaScript-initiated are not exposed to their security monitor.

Guha \etal{} also used finite state automata to encode web security policies~\cite{GuhaKJ09}. Their approach is based on three steps: first, they apply a static analysis for JavaScript to construct the control flow graph of an Ajax application to protect and then they use it to synthesize a request graph, which summarizes the expected input/output behavior of the application. Finally, they use the request graph to instruct a server-side proxy, which performs a dynamic monitoring of browser requests to prevent observable violations to the expected control flow. The security enforcement can thus be seen as the computation of a finite state automaton built from the request graph. Their technique, however, is only limited to Ajax applications and operates at the server side, rather than at the browser side.

\subsection{Browser-Side Defenses}
The present paper positions itself in the popular research line of extending web browsers with stronger security policies. To the best of our knowledge, this is the first work which explicitly focuses on web protocols, but a number of other proposals on browser-side security are worth mentioning. Enforcing information flow policies in web browsers is a hot topic nowadays and a few fairly sophisticated proposals have been published as of now~\cite{GroefDNP12,HedinBS15,BichhawatRGH14,RajaniB0015,BauerCJPST15}. Information flow control can be used to provide confidentiality and integrity guarantees for browser-controlled data, but it cannot be directly used to detect deviations from expected web protocol executions, which instead are naturally captured by security automata. Combining our approach with browser-based information flow control can improve its practicality, because a more precise information flow tracking would certainly help a more permissive security enforcement.

A number of browser changes and extensions have been proposed to improve web session security, both from the industry and the academia. Widely deployed industrial proposals include Content Security Policy (CSP) and HTTP Strict Transport Security (HSTS). Notable proposals from the academia include Allowed Referrer Lists~\cite{CzeskisMKW13}, SessionShield~\cite{NikiforakisMYJJ11}, Zan~\cite{TangDK11}, CSFire~\cite{RyckDJP11}, Serene~\cite{RyckNDPJ12}, CookiExt~\cite{BugliesiCFK15}, SessInt~\cite{BugliesiCFKT14} and Michrome~\cite{CalzavaraFGM16}. Moreover, JavaScript security policies are a very popular research line in their own right: we refer to the survey by Bielova~\cite{Bielova13} for a good overview of existing techniques. None of these works, however, tackles web protocols.

\section{Conclusion}
We presented \WPE, the first browser-side security monitor designed to address the security challenges of web protocols, and we showed that the security policies enforceable by \WPE{} suffice to prevent a large number of real-world attacks. Our work encompasses  a thorough review of well-known attacks reported in the literature and an extensive experimental analysis performed in the wild, which exposed several undocumented security vulnerabilities fixable by \WPE{} in existing OAuth 2.0 implementations. We also discovered a new attack on the Google implementation of SAML 2.0 by formalizing its specification in \WPE. In terms of compatibility, we showed that \WPE{} works flawlessly on many existing websites, with the few compatibility issues being caused by custom implementations deviating from the OAuth 2.0 specification, one of which introducing a critical vulnerability. 
In the end, we conclude that the browser-side security monitoring of web protocols is both useful for security and feasible in practice.

As to future work, we observe that our current assessment of \WPE{} in the wild only covers two specific classes of vulnerabilities, which can be discovered just by navigating the tested websites: extending the analysis to cover active attacks (in an ethical manner) is an interesting direction to get a better picture of the current state of the OAuth 2.0 deployment. We would also like to improve the usability of \WPE{} by implementing a more graceful error handling procedure: \eg, when an error occurs, we could give users the possibility to proceed just as it routinely happens with invalid HTTPS certificates. Using more descriptive warning messages may also be useful for web developers that are visiting their websites with \WPE{} so that they can understand the issue and provide the appropriate fixes to the server side code. Finally, we plan to identify automated techniques to synthesize protocol specifications for \WPE{} starting from observable browser behaviours in order to make it easier to adopt our security monitor in an industrial setting. 

\medskip
\noindent
\textbf{Acknowledgments.} This work has been partially supported by the European Research Council (ERC) under the European Union’s Horizon 2020 research (grant agreement No 771527-BROWSEC), by Netidee through the project EtherTrust (grant agreement 2158),  by the Austrian Research Promotion Agency through the Bridge-1 project PR4DLT (grant agreement 13808694) and COMET K1 SBA. The paper also acknowledges support from the MIUR project ADAPT and by CINI Cybersecurity National Laboratory within the project FilieraSicura: Securing the Supply Chain of Domestic Critical Infrastructures from Cyber Attacks funded by CISCO Systems Inc. and Leonardo SpA.

%\balance
\bibliographystyle{abbrv}
\bibliography{main}

\appendix

\section{Sample XML Specification}
\label{sec:xml}
% !TeX root = main.tex
Figure~\ref{fig:oauth-no-state} shows the XML specification of the OAuth 2.0 automaton in Figure~\ref{fig:oauth-dfa}. The protocol is enclosed within \texttt{<Protocol>} tags and describes the flow as a sequence of requests and responses. For every message we detail its pattern, possibly specifying the endpoint and a list of parameters for requests or a list of headers for responses.

Identifiers can be introduced in the protocol flow specification by adding the \texttt{id} attribute to the tag of the message component of interest. Additional identifiers can be defined within \texttt{<Definition>} tags, where the value that is associated to the new identifier is the part of the \texttt{<Source>} matching the regular expression \texttt{<Regexp>}. If the regular expression contains a capturing group, denoted by parenthesis, only the string matching the group is selected. The syntax \texttt{\$\{id\}} can be used to refer to the value bound to the identifier \texttt{id}.

Security policies are defined within \texttt{<Secrecy>} and \texttt{<Integrity>} tags. The secrecy policy specifies that the value in \texttt{<Target>} must be sent only to the enumerated origins. The integrity policy specifies that the value in \texttt{<Target>} must match the content of \texttt{<Matches>}, which can possibly be a regular expression.

\begin{figure*}[ht]
\footnotesize
\begin{lstlisting}[xleftmargin=5.0ex, basicstyle=\ttfamily, columns=fullflexible,keepspaces=true, numbers=left, numberstyle=\tiny, escapechar=@]
<Specification name="google-explicit-nostate">
    <Protocol>
        <Request method="GET" desc="req_init">
            <Endpoint>
                <Regexp> https://accounts\.google\.com/o/oauth2/(?:.*?/)?auth </Regexp>
            </Endpoint>
            <Parameter name="response_type"> code </Parameter>
            <Parameter name="redirect_uri" id="req_init_redirect_uri" />
        </Request>
        <Response desc="resp_init">
            <Endpoint>
                <Regexp> https://accounts\.google\.com/o/oauth2/(?:.*?/)?auth </Regexp>
            </Endpoint>
            <Header name="Location" id="resp_init_location" />
        </Response>
        <Request method="GET" desc="req_code">
            <Endpoint id="uri2"/>
            <Parameter name="code">
                <Regexp> [^\s]{40,} </Regexp>
            </Parameter>
        </Request>
    </Protocol>
    <Identifiers>
        <Definition id="uri1">
            <Source> ${req_init_redirect_uri} </Source>
            <Regexp> ^(https?://.*?)(?:\?|$) </Regexp>
        </Definition>
        <Definition id="origin">
            <Source> ${req_init_redirect_uri} </Source>
            <Regexp> ^(https?://.*?/).* </Regexp>
        </Definition>
        <Definition id="authcode">
            <Source> ${resp_init_location} </Source>
            <Regexp> [?&amp;]code=(.*?)(?:&amp;|$) </Regexp>
        </Definition>
    </Identifiers>
    <Policy>
        <Secrecy> <!-- the auth code contained in the Location header must be kept secret -->
            <Target> ${authcode} </Target>
            <Origin> ${origin} </Origin>
            <Origin> https://accounts.google.com/ </Origin>
        </Secrecy>
        <Integrity> <!-- the last message must be sent to the redirect URI initially specified -->
            <Target> ${uri2} </Target>
            <Matches> ${uri1} </Matches>
        </Integrity>
    </Policy>
</Specification>
\end{lstlisting}
\caption{XML specification for the automaton in Figure~\ref{fig:oauth-dfa}.}
\label{fig:oauth-no-state}
\end{figure*}

\end{document}